\documentclass[review]{elsarticle}

\usepackage{algorithmic}
\usepackage{graphicx}
\usepackage{amssymb}
\usepackage{lipsum}
\usepackage{ifpdf}
\usepackage{array}

\usepackage{amsthm}
\usepackage{mathtools}
\usepackage{color}
\usepackage[pdftex,dvipsnames]{xcolor}  
\usepackage{xargs}                      
\usepackage[colorinlistoftodos,prependcaption,textsize=tiny]{todonotes}
\usepackage{url}
\DeclareMathOperator*{\argmax}{argmax} 

\usepackage{float}
\usepackage[]{algorithm2e}
\usepackage{booktabs} 
\usepackage{soul}
\usepackage[section]{placeins}
\newtheorem{theorem}{Theorem}

\newtheorem{lemma}[theorem]{Lemma}

\journal{Engineering Applications of Artificial Intelligence}

\begin{document}

\begin{frontmatter}

\title{Real-Time EEG Classification via Coresets for BCI Applications}


\author[EitanAndDanny]{Eitan Netzer}

\author[Alex_Addr]{Alex Frid\corref{mycorrespondingauthor}}
\cortext[mycorrespondingauthor]{Corresponding author}
\ead{alex.frid@gmail.com}

\author[EitanAndDanny]{Dan~Feldman}

\address[EitanAndDanny]{Robotics and Big Data Lab, Computer Science department, University of Haifa, Haifa, Israel.}
\address[Alex_Addr]{Laboratory of Clinical Neurophysiology, Faculty of Medicine, Technion IIT, Haifa, Israel.}

\begin{abstract}
	A brain-computer interface (BCI) based on the motor imagery (MI) paradigm translates one's motor intention into a control signal by classifying the Electroencephalogram (EEG) signal of different tasks. However, most existing systems either (i) use a high-quality algorithm to train the data off-line and run only classification in real-time, since the off-line algorithm is too slow, or (ii) use low-quality heuristics that are sufficiently fast for real-time training but introduces relatively large classification error. 
	
	In this work, we propose a novel processing pipeline that allows real-time and parallel learning of EEG signals using high-quality but possibly inefficient algorithms. This is done by forging a link between BCI and core-sets, a technique that originated in computational geometry for handling streaming data via data summarization. 
	
	
	We suggest an algorithm that maintains the representation such coreset tailored to handle the EEG signal which enables: (i) real time and continuous computation of the Common Spatial Pattern (CSP) feature extraction method on a coreset representation of the signal (instead on the signal itself) , (ii) improvement of the CSP algorithm efficiency with provable guarantees by applying CSP algorithm on the coreset, and (iii) real time addition of the data trials (EEG data windows) to the coreset.

	
	For simplicity, we focus on the CSP algorithm, which is a classic algorithm. Nevertheless, we expect that our coreset will be extended to other algorithms in future papers. In the experimental results we show that our system can indeed learn EEG signals in real-time for example a 64 channels setup with hundreds of time samples per second. Full open source is provided to reproduce the experiment and in the hope that it will be used and extended to more coresets and BCI applications in the future.
\end{abstract}


\end{frontmatter}


	



\section{Introduction}
\label{S:1}
Brain-computer interfaces (BCI's) translate brain signals into a control signal without using one's actual movements or peripheral nerves. BCI's based on Electroencephalogram (EEG) recordings have many advantages, such as short time constraints, less environmental limits, and the requirement of relatively inexpensive equipment. On the other hand, EEG introduces a high amount of noise and requires handling a large amount of data in real-time. In addition, those systems usually require a time consuming training phase.

In recent years many techniques were developed for the EEG-MI based BCI systems which introduced high classification accuracy. For example, the average accuracy of classifying imaginary left and right hand movement in some cases can achieve more than 90\% \cite{ramoser2000optimal,dornhege2004boosting}. Many of these techniques are based on Common Spatial Patterns (CSP) signal decomposition \cite{ang2008filter, koles1990spatial, dornhege2006optimizing, pfurtscheller1999eeg, fukunaga2013introduction}. Nevertheless, these systems still have very limited usage in real-life applications. This is partially because the algorithms used in those systems focus on analysing multi-channelled and densely sampled  EEG signal, which requires relatively expansive equipment due to the need of high processing power and memory.

An example of such computational bottlenecks in these systems is the CSP algorithm \cite{blankertz2008optimizing}. In essence it is a "batch processing algorithm", i.e. whenever a new sample is introduced, the algorithm needs to be re-trained in order to find the updated spatial filters.

\begin{figure*}[!t]
	\center
	\includegraphics[scale=0.40]{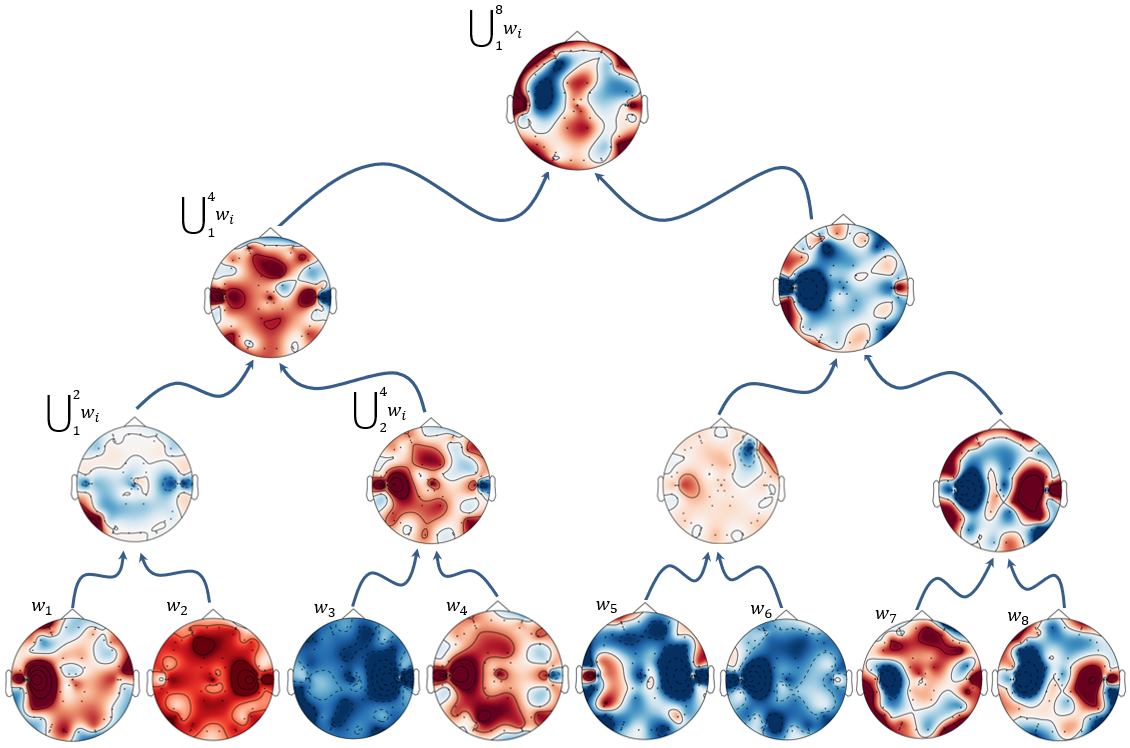}
	\caption{CSP weights representation using Coresets. The leafs of the tree can represent CSP algorithm result (i.e. weights) computed on individual sample, a window or a single EEG trial. The verteces of the tree represent a "coreset CSP" computed on all the samples until that point (a unification of coresets in proposed algorithm opposed to recomputation of the CSP algorithm).}
	\label{fig:scheme2}
\end{figure*}

In this work, we present a method that is based on coreset representation \cite{AgaHarVar04} of the EEG signal that can be executed prior to the CSP signal decomposition (see visualization in figure \ref{fig:scheme2} ), and thus can reduce both the computational cost and memory consumption without losing classification accuracy. Which in turn will allow to use cheaper and low powered hardware in BCI devices.

\subsection{Background}
The method of common spatial pattern was first used in EEG analysis to extract abnormal components from the clinical EEG \cite{koles1991quantitative} and in later stages, it was adopted to BCI applications. In essence, this method weights each electrode according to its importance for the discrimination task and suppresses noise in individual channels by using correlations between neighboring electrodes. Let \(X_{1}\in\mathbb{R}^{d\times t_{1}}\), and \(X_{2}\in\mathbb{R}^{d\times t_{2}}\) be multivariate signals of degree $d$, where \(d\) is the number electrodes or sensors and \(t_{i}\) is the number of time samples. For example in MI \(X_{1}\), and \(X_{2}\) may represent the signal associated with subject imagining of moving his left or right hand. CSP determines for every $w\in\mathbb{R}^d$, such that non zero ${\left\Vert  wX_{2} \right\Vert ^{2}}$, the component \(w^{T}\) that maximizes the ratio of variance between \(X_{1}\) and \(X_{2}\) [5]:

\begin{equation} \label{eq:oldW}
w\in \underset{w}{\arg\max}\frac{\left\Vert wX_{1} \right\Vert ^{2}}{\left\Vert  wX_{2} \right\Vert ^{2}}
\end{equation}

In order to solve the aforementioned problem\cite{legendre1989spatial}, the CSP algorithm first computes the covariance matrices \(R_{i}=\frac{X_{i}X_{i}^{T}}{t_{i}}\) where $i ={1,2}$, then simultaneous diagonalization of both matrices  ($R_{2}^{-1}R_{1}$) using generalized eigenvalue decomposition is performed. Since $X_i$ is of degree $d$ $R_i$ is of full degree and invertible. Let \(U\) be eigenvectors matrix and, \(D\) a diagonal matrix of eigenvalues \(\left\{ \lambda_{1},\lambda_{2},...,\lambda_{d}\right\} \) in decreasing order, such that \(U^{-1}R_{1}U=D\) and \(U^{-1}R_{2}U=I_{d}\) where \(I_{d}\) is the identity matrix. \(R_{2}^{-1}R_{1}=UU^{-1}UDU^{-1}=UDU^{-1}\) hence its equivalent to the eigendecomposition of \(R_{2}^{-1}R_{1}\),  and \(w^{T}\) correspond to the first column of  \(U\). A more detailed description provided in the Algorithm 1 below.

\begin{algorithm}
	\KwData{\(X_{1}\in\mathbb{R}^{d\times t_{1}}\) and \(X_{2}\in\mathbb{R}^{d\times t_{2}}\) be multivariate signals, where \(d\) is the number of electrodes or sensors and \(t_{i}\) is the number of time samples from class $i={1,2}$}
	\KwResult{component \(w^{T}\) that maximizes the ratio between \(X_{1}\) and \(X_{2}\); as in equation (\ref{eq:oldW}).}
	\textbf{Function} CSP($X_1$,$X_2$)
	\begin{enumerate}
		
		\item \(R_{1}\leftarrow\frac{X_{1}X_{1}^{T}}{t_{1}}\), \(R_{2}\leftarrow\frac{X_{2}X_{2}^{T}}{t_{2}}\)
		\item \(A\gets R_{2}^{-1}R_{1}\)
		\item compute \(D,U \gets \) the  eigendecomposition of \(A\)
		\item \(w^{T} \gets U_{1}\) (the first column of \(U\))
		\item return \(w\) \newline
	\end{enumerate}
	
	\caption{Computation of Common Spatial Pattern (CSP)}
\end{algorithm}

The main drawback for using this algorithm in real time applications lays in its time and space complexity. Indeed computing the covariance matrices (step 1 in the Algorithm 1) is an \(O\left(d^{2}\left(t_{1}+t_{2}\right)\right)\) time complexity, followed by inverting a \(d\times d\) matrix which takes \(O\left(d^{3}\right)\) time complexity and then finding eigenvalues, and eigenvectors with \(O\left(d^{2}\right)\). The total time complexity is thus \(O\left(d^{2}\left(t_{1}+t_{2}+d\right)\right)\), and require space (memory) complexity of \(O\left(d\left(t_{1}+t_{2}\right)\right)\), when typically \(d<<t_{i}\) (due to the EEG's high sampling rate and it's continuous operation). This dependency on time samples eliminates the possibility of using CSP in real time streaming. Our coreset based algorithm has a fixed time and space complexity of \(O\left(d^{2}\right)\), allowing real time streaming applications per new added sample.
\begin{center} 
	\begin{table} \caption{Complexity bound, Coreset versus Traditional CSP}
		\label{tbl:coresetVsTra}
		\center
		\begin{tabular}{|c || c | c||}
			\hline
			Algorithm / Complexity & Time  & Space \\ [0.5ex]\hline
			Traditional CSP & \(O\left(d^{2}\left(t_{1}+t_{2}+d\right)\right)\) & \(O\left(d\left(t_{1}+t_{2}\right)\right)\) \\ \hline
			Coreset CSP  & \(O\left(d^{2}\right)\) & \(O\left(d^{2}\right)\) \\\hline
		\end{tabular}
	\end{table}
\end{center}

Past attempts have been made to reduce this computational cost. For example, \cite{zhao2008incremental} proposed an incremental way to update the spatial filters where new sample introduced to the algorithm, i.e. the feature extraction process performed is in on-line fashion. 
In \cite{ross2008incremental} it was proposed to preform incremental learning algorithm that is based on incremental algorithms for principal component analysis with a forgetting factor. This is essentially an adaptation of the algorithm presented in \cite{levey2000sequential}. Suppose $X\in\mathbb{R}^{d\times t}$ represent a set of sample points, where $d$ is the number of features and $t$ is the number of samples  and $q\in{Q}$ a quire from a set queries or family of models to optimize equation (\ref{eq:oldW}). 
\\



Nevertheless, those methods have one or all of the following disadvantages: 
\begin{enumerate}
	\item The \textbf{Running time} that it takes to minimize $f(X,q)$ might be impractical. A possible solution is to use faster heuristics with no provable approximations, but the cost might be a weaker classifier. In the context of EEG, in many applications the signal from the brain received in \emph{real-time} and the model must be updated in a fraction of a second.
	\item  \textbf{Memory} management issues arise for large signals that cannot fit into memory (RAM), or can fit into memory but are too large to be processed by the optimization algorithm. In the context of EEG, the memory (signal's length) increases because of the many channels, the frequency of the sampling, and possibly the number of users.
	\item \textbf{On-line} on the-fly data points that are received from the signals are classified but are not being used to update the model (i.e. learning) to improve classification over time. In the context of EEG, we might get feedback from the user or the real-world regarding the last classification and we wish to use the new information for next samples.
	\item \textbf{Parallel computation.} Even if the algorithm is sufficiently enough, it may not be clear on distributed data how to run it in parallel over few parallel computation threads to reduce its running time and take the advantage of the computation power that can be used by modern multi-core CPU or GPUs. In the context of EEG, the input is also parallel when it is received from either few users or few BCI channels of the same user.
	\item \textbf{Distributed computation. }Even if the algorithm supports parallel computation, it may not support distributed computation. Here, the input signal itself is partitioned between different machines (cloud, device) or threads as in GPUs, that have no shared memory and little communication data between them might be expensive. In the context of EEG, each user might be connected to a different computation device, but we aim for a single classifier. Similarly, when the signal is streamed to a cloud of machines, we need parallel computation with  little shared memory via network communication which is relatively expensive and slow.
	\item \textbf{Dynamic data. }Even on-line or parallel algorithms usually are not able to handle deletion of samples in the signals. In the context of EEG, this is the case when we use the \emph{sliding window} model. Here, we wish the classifier to represent only samples from the last $t$ seconds. That is, every time that a new sample point arrives, the sample that was received $t$ seconds ago should be deleted. The classifier is then the one that minimizes $f(X,q)$ above where $X$ is only the set of remaining samples.
	\item \textbf{Handling Variations and Constrains.} In practice and industry we usually have constraints, business rules and expert rules that are based on very specific scenarios, signals, laws, users or applications. For example, we want to minimize $f(X,q)+\left\lVert q \right\rVert$ instead of $f(X,q)$, where $\left\lVert q \right\rVert$ is called a regularization term that is used to obtain a simpler classifier, or we want to minimize $f(q)$ but under some specific constraints where $q\in Q'\subset Q$.
\end{enumerate}


In this work, we present an alternative approach for optimizing the CSP algorithm learning and its variants by using coreset representation of the data. This in turn satisfies the aforementioned requirements, i.e. allows on-line learning, constant memory consumption, computational efficiency, parallel and distributed computation, while allowing not only addition but also a deletion of the data.
The rest of the paper is organized as follows: in Section 2, an introduction to coresets and a formulation of the corsets for EEG is provided. Section 3 presents the proposed algorithm and it's analysis and theorems. Section 4 compares the practical performance of coreset-based CSP algorithm with a traditional algorithm on well known EEG dataset. Finally the last section, Section 5, summarizes and concludes the our work. 

\section{Related work: Coresets for EEG real-time  processing}
The term coreset was coined by Agarwal, Har-Peled and Varadarajan in~\cite{AgaHarVar04}.
First, coresets improved the running of many open problems in \textbf{computational geometry} (e.g.~\cite{aga1,aga2,HP01,feldman2007ptas}); See surveys in~\cite{sur,CzuSoh07a,Phillips16}. Later, coresets were designed for obtaining the first PTAS or LTAS (polynomial/linear time approximation schemes) for more classic and graph problems in \textbf{theoretical computer science}~\cite{FraSoh05,czumaj2005approximating,FrahlingIS08,BuriolFLS07}, and more recently under the name "composable coresets"~\cite{indyk2014composable,mirrokni2015randomized,aghamolaei2015diversity}. Coresets are usually used when we want to approximate large data by a simple model with relatively few parameters, and are used less for real-time systems as in this paper. In particular, in projective clustering~\cite{feldman2013turning,aga1,santosh,HV02,ProcAga02,Boug,agakmean,DeshpandeEtAl06,varadarajan2012sensitivity} the model is a set of $k$ points, lines or subspaces, with an appropriate fitting cost. This is also a common setting in \textbf{machine learning}~\cite{feldmanmik,feldman2011scalable,tsang2005core,lucic2016strong,bachem2016approximate,lucic2015tradeoffs,bachem2015coresets,huggins2016coresets,rosman2014coresets,reddi2015communication}. More applied research was suggested e.g. by Rus et al.~\cite{feldmanmik,sung2012trajectory,feldman2013idiary,rusprivate,rosman2014coresets} Krause~\cite{feldman2011scalable,bachem2015coresets,lucic2016strong,bachem2016approximate} , Smola~\cite{reddi2015communication} or Sochen~\cite{feigin2011high,feldman2013learning,alexandroni2016coresets} in \textbf{image processing} with applications for medicine.

Improved techniques for using coresets for distributed data and low communication on the cloud, with both theoretical guarantees and experimental results were recently suggested in \textbf{data mining} conferences such as~\cite{FT15,liberty2013simple}. Classical \textbf{optimization} techniques such as Frank-Wolfe~\cite{clarkson2010coresets} and semi-definite programming~\cite{cohen2016geometric} appear to produce deterministic and smaller types of coresets. In \textbf{Numerical linear algebra} coresets were suggested for matrix approximations~\cite{drineas2006fast,DrineasMM06,DasguptaDHKM08} using random projections, called \emph{sketches}.
The first coresets for \textbf{signal processing} with applications to GPS or video data were suggested in~\cite{feldman2013idiary,rosman2014coresets,sung2012trajectory}. The first results for \textbf{probabilistic databases} appeared recently~\cite{munteanu2014smallest,DBLP:conf/esa/HuangLPW16}

In this work we show that coreset paradigm can improve the computations described different sections, including provable guarantees of complexity in terms of training/inference time and memory, also for EEG applications. In particular, we demonstrate how coresets can be used to train a classifier in real-time for EEG signals.
More details on coresets and the theoretical proofs for the computation models below can be found e.g. in~\cite{sariela,sur,FL11,FT15}.

As in the previous section, consider the problem of minimizing $f(X,q)$ over $q\in Q$, where $Q$ is a set of query and $f$ is a function $f(X,q)\rightarrow \mathbb{R^+}$. In this paper a coreset for this optimization problem, as in the coreset for CSP, would be another set $C$ such that $f(C,q)=f(X,q)$ for every $q\in Q$.  


The fact that the coreset approximates the original data in the above sense is not sufficient to handle the computation models in the previous sections. What we need for these is a \emph{composable coresets} construction. This means that the coreset construction satisfies two properties. First, the union of two coresets is a coreset. That is coresets are mergable in the sense that if $f(C_a,q)$ approximates $f(X_a,q)$ and $f(C_b,q)$ approximates $f(X_b,q)$ then $f(C_a\cup C_b,q)$ approximates $f(X_a\cup X_b,q)$, for every $q\in Q$. The second property is that we can compress a pair of coresets to obtain another coreset. Formally, we can compute a coreset $C'$ such that $f(C',q)$ approximates $f(C_a\cup C_b,q)$ for every $q\in Q$. 
Using these construction properties, we can build a coreset tree (see \ref{fig:scheme} for an example of such tree).



\paragraph{Running time} Let $s\geq1 $ be an integer so that if the input set $X$ to the coreset construction algorithm is of size $|X|\leq s$, then the resulting coreset is of cardinality $|C|\leq s/2$. Assume that this construction takes $g(s)$ time. 
As is shown in Fig.~\ref{fig:scheme2}, we can now merge-and-reduce a data set $X$ of size $n$ recursively to obtain its coreset under the above models. That is, we partition the input signal into subsets of samples, each of size $s/2$ (the leaves of the binary tree). In the next level of the tree we take every union of coresets in a pair of leaves (that consists of $s$ points) and reduce them back to $s$ points. This takes overall $(2n/s)\cdot g(s)$ time for the $n/s$ leaves, which is \emph{linear} in $n$ even if our coreset constructions takes, say $g(n)=n^{10}$ time for input of $|X|=n$ points.

\paragraph{Streaming data} Handling streaming data can be done in a similar way, where the leaves arrive on-the-fly. Every set in a pair of leaves is reduced to a single coreset and the pair of leaves are then deleted. Hence, there are no more than one pair of leaves during the streaming. Similarly, whenever there are two inner nodes in a level of the tree we reduce them to a single node in the higher level. At any given moment there is at most a single coreset in each of the $O(\log n)$ levels of the tree for the $n$ points seen so far in the stream.

\paragraph{Distributed data} When the data is both streamed and distributed, say, to $M=2$ machines, we assume that every second point is being sent to the second machine, and the rest (odd) points are being sent to the first machine. This can be done directly from the users, or from a main server. Each machine is then independently computing the merge-and-reduce tree of its points, as explained in the previous paragraph. The speed of computation and streaming then grows linearly with $M$. This is known as ``embarrassingly parallel" independent computation. When a coreset for the complete data is needed, a main server can collect the coreset of each tree on each machine. This requires communication to a main server, however, since each machine sends only the coreset of its data, only $O(s)$ bits are sent in parallel from the $M$ machines.

\paragraph{Dynamic computations} To support deletion of input points (as in the sliding window model above), we need to store the complete merge-and-reduce tree as a balanced 2-3 binary tree whose leaves are the input points (a single point to a leaf). Here, every inner node, which is a root of a sub-tree, contains the coreset for the leaves in this sub-tree. When a point (leaf) is being deleted, we only need to update its $O(\log n)$ ancestors in the tree with their corresponding coresets. Recomputing these coresets takes $f(s)\cdot O(\log n)$ time per point insertion/deletion, which is only logarithmic in $n$, the number of points seen so far.


\paragraph{Real Time Training as a feedback} Our coreset allows real time training of the model. Brain related signals such as EEG are generated by a user. Using the system reaction of an updated (a.k.a inference) allows that not only the system will "learn" the human participator but the user will learn the system. This allows the user to aim his thought best such that to control the system, shortening and updating training phase. 

\section{Proposed Algorithm}

\begin{figure*}[!t]
	\center
	\includegraphics[width=\textwidth,height=\textheight,keepaspectratio]{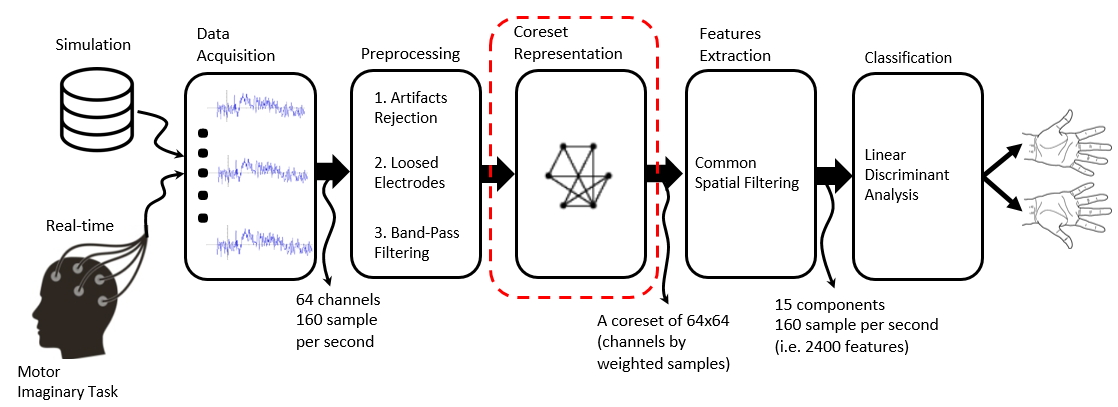}
	\caption{Proposed system diagram. The additional process of using coresets to compress the EEG data is marked by red box.}
	\label{fig:scheme}
\end{figure*}

A full description of the algorithm flow is provided in this section, along with its graphical illustration (see Fig.~\ref{fig:scheme}). As can be seen in Fig. \ref{fig:scheme}, the input signal is streamed out either from the database (i.e. a real-time simulation) or from  EEG headset. After acquiring the data, the EEG signal undergoes preprocessing stage, during which (i) various signal artifacts are checked (such as eye blinking and loosed electrodes) and then (ii) the signal is band-passed to frequencies containing MI information. In the next step, a coreset is fitted to the EEG data, which leads to a more compact representation of the signal. 

On this compact representation, the CSP algorithm is applied to find the discriminative spatial filters. Next, the last step of the algorithm is the classification of the MI task type (i.e. left or right hand movement). This step is performed  using Linear Discriminant Analysis (LDA) algorithm.

Each of these steps is described in detail below:






The EEG signal \(X_{i,t}\) at time \(t\) of each class \(i\in\left\{ 0,1\right\} \) is represented using a coreset in the following way: 

For each new time sample (or a processing window), the coreset compress the signal to be bounded by the number of electrode leads (i.e. sensors). When a new time sample is entered into the system, the current signal is represented by \(d\times\left(d+1\right)\) matrix. In order to compress it back to \(d\times\left(d+1\right)\), first an SVD matrix decomposition is applied resulting in \(U,S,V=svd\left(X_{i,t}\right)\) where \(S\) is a diagonal eigenvalue matrix and \(U,V\) are matrices whose columns are the singular vectors. 

 Let \(Y=US\), then \(\left\Vert X\right\Vert ^{2}=\left\Vert YV\right\Vert ^{2}=\left(YV\right)\left(YV\right)^{T}=YVV^{T}Y^{T}=\left\Vert U\cdot S\right\Vert ^{2}\). Then, we update the coreset representation of the signal to be \(C_{i,t}=U\cdot S\), an eigenvalue \(d\times d\) matrix, and repeat this procedure for each incoming sample (or a processing window). The time complexity of adding a sample to the coreset is \(O\left(d^{3}\right)\) and space complexity is \(O\left(d^{2}\right)\), when typically \(d\) is very small. See Algorithm \ref{alg:EEG_Signal_Coreset}.

\begin{algorithm}
	\KwData {\(C\) empty matrix or \(C\in\mathbb{R}^{d\times d}\) coreset reprsention.\(x\in\mathbb{R}^{d}\)  a multivariate time sample of a signal, where \(d\) is the number of electrodes or sensors.}
	\KwResult{An update by sample coreset representation of the  input signal. }
	\textbf{Function} EEGSignalCorest($C,x$):\newline
	\begin{enumerate}
		\item  \(C\leftarrow C || x\)
		\item  if  \(Row(C)>d\):
		\item \hspace{0.5cm} \(U,S,V=svd(self.C)\)
		\item \hspace{0.5cm} \(C\leftarrow U\cdot S\) 
		\item Return \(C\)
		\newline
		\caption{EEG signal coreset construction}
		\label{alg:EEG_Signal_Coreset}
	\end{enumerate}
\end{algorithm}
\FloatBarrier
Additionally we show that it is possible to concatenate a window of samples to a coreset. \\

\FloatBarrier

\begin{lemma} For every $ w\in\mathbb{R}^d:\\
	\begin{Vmatrix} w 
	\begin{bmatrix} 
	U\cdot{S} \\
	x_{n+1}
	\end{bmatrix}
	\end{Vmatrix} ^2
	=
	\begin{Vmatrix} w 
	\begin{bmatrix} 
	U\cdot{S} \\
	x_{n+1}
	\end{bmatrix}
	\begin{bmatrix} 
	V & 0 \\
	0 & 1
	\end{bmatrix}
	\end{Vmatrix} ^2
	=
	\begin{Vmatrix} w 
	\begin{bmatrix} 
	U\cdot{S}\cdot{V} \\
	x_{n+1}
	\end{bmatrix}
	\end{Vmatrix} ^2
	=
	\begin{Vmatrix} w 
	\begin{bmatrix} 
	X_{1,2,...,n} \\
	x_{n+1}
	\end{bmatrix}
	\end{Vmatrix} ^2
	$, where $\begin{bmatrix} 
	X_{1,2,...,n} \\
	x_{n+1}
	\end{bmatrix}$ is the concatenation of matrix $X_{1,2,...,n}\in\mathbb{R}^{d \times n }$ of samples $1$ to $n$ with the vector $x_{n+1}$ of sample $n+1$, $S\in \mathbb{R}^{d \times d } $ diagonal matrix of the eigenvalues, $V\in \mathbb{R}^{d \times d }$ matrix of eigenvectors, and $\begin{bmatrix} 
	U\cdot{S} \\
	x_{n+1}
	\end{bmatrix}$ is the concatenation of samples $1$ to $n$ after svd decomposition with with the vector $x_{n+1}$ of sample $n+1$.
\end{lemma}

\subsection{Common Spatial Patterns and Corsets}

\hspace*{0.5cm} The coreset signal \(C_{i,t}\) for both signal \(i=1,2\), is used in every new sample to maximize the following equation
\begin{equation} \label{eq:newCSP}
w=\argmax_w\frac{\left\Vert wC_{1,t}\right\Vert ^{2}}{\left\Vert wC_{2,t}\right\Vert ^{2}}
\end{equation}
where $w$ is equivalent of Eq.~\eqref{eq:oldW} for the real-time process. Using the coreset signal, we are able to compute the CSP component using fewer samples and much faster. For diagonal matrix and real unitary matrix, \[R_{i}=\left(U_{i}S_{i}V_{i}\right)\left(U_{i}S_{i}V_{i}\right)^{T}=U_{i}S_{i}V_{i}V_{i}^{T}S_{i}^{T}U_{i}^{T}=U_{i}S_{i}^{2}U_{i}^{T}\]the covariance matrix \(R_{1}=U_{1}\cdot S_{1}^{2}\cdot U_{1}^{T}\), where for \(S_{1}^{2}\) we need calculate for the main diagonal since \(S_{1}\)  is a diagonal matrix, \(R_{2}^{-1}=\left(U_{2}\cdot S_{2}^{2}\cdot U_{2}^{T}\right)^{-2}=U_{2}^{T}\cdot S_{2}^{-2}\cdot U_{2}\), again for
\(S_{2}^{-2}\) we just calculate for the main diagonal since \(S_{2}\) is a diagonal matrix and \(V_{2}\) is unitary and real matrix hence \(U_{2}^{-1}=U_{2}^{T}\). The problem is reduced to computing:
\[\left\Vert R_{2}^{-1}R_{1}\right\Vert ^{2}=\left\Vert U_{2}^{T}S_{2}^{-2}U_{2}^{T}U_{1}S_{1}^{2}U_{1}^{T}\right\Vert ^{2}\]

The complexity bound of adding a sample is determined by Algorithm \ref{alg:EEG_Signal_Coreset}. The time complexity of calculating the covariance matrix and inverting is \(O\left(d^{2}\right)\) because \(S_{i}\) is diagonal matrix and \(U_{i}\)
is unitary and real matrix. Finding the eigenvalues and eigenvectors takes time \(O\left(d^{3}\right)\), resulting with time complexity of \(O\left(d^{3}\right)
\) and space complexity of \(O\left(d^{2}\right)\). See Algorithm \ref{alg:CSPCoreset} for additional details. Figure \ref{fig:scheme2} depicts graphically the advantage of using coresets for CSP computation. Each leaf in the graph represents a spatial filter (i.e. "$w$") computed by the CSP algorithm. 
\\
When an additional trial is presented to the system, the traditional CSP computation will require re-computation of all the previous EEG trials along with the new one. In comparison using the "coreset based CSP" that will only require a unification of two coresets. This is comparison to the traditional approach that will require re-computation of all the data. Additionally, this coreset representation allows parallel computation and removal of trials or even a group of trials from the CSP filters without recomputing the CSP algorithm on the remaining data.

\begin{algorithm}
	\KwData {\(C_1\in\mathbb{R}^{d\times d}\) a coreset of $X_1$\newline
		\(C_2\in\mathbb{R}^{d\times d}\) a coreset of $X_2$\newline
		\(x\in\mathbb{R}^{d}\)  a multivariate time sample of a signal, where \(d\) is the number electrodes or sensors. \newline
		\(y\in\left\{ 0,1\right\} \) the label of current sample}
	\KwResult{component \(w^{T}\) that maximize the ratio between \(X_{1}\) and \(X_{2}\) see equation \ref{eq:newCSP}.}
	
	\textbf{Function} ComputeW($C_1$,$C_2$,x,y):
	
	\eIf{y==1}{
		$C_1 \leftarrow EEGSiganlCoreset(C_1,x)$
	}{
		$C_2 \leftarrow EEGSiganlCoreset(C_2,x)$
	}
	w = CSP($C_1$,$C_2$)\\
	\textbf{Return}  w 
	
	\caption{CSP coreset representation}
	\label{alg:CSPCoreset}
\end{algorithm}


\subsection{Classification Scheme}

A common classifier for BCI and MI task is the Linear Discriminant Analysis (LDA) method \cite{pfurtscheller2000current}. LDA is a generalization of the "classical" Fisher's linear discriminant frequently used in statistical pattern recognition for finding a linear combination of features that separates between classes \cite{fisher1936use, martinez2001pca}. If we let \(X\) be a feature vector and \(y\) a known class label, LDA assumes that the conditional probability density functions \(p\left(x\mid y=2\right)\) and \(p\left(x\mid y=1\right)\) are normally distributed with mean and covariance parameters \(\left(\mu_{1},\varSigma\right)\) and \(\left(\mu_{2},\varSigma\right)\), respectively where \(\varSigma\) is Hermitian and invertible. It predicts by using Bayes optimal solution with threshold \(T\) with log likelihood ratio, \(\varSigma^{-1}\left(\mu_{2}-\mu_{1}\right)x>C\) where \(C\) is a constant s.t. \(C=\frac{1}{2}\left(T-\mu_{1}^{T}\varSigma^{-1}\mu_{1}+\mu_{2}^{T}\varSigma^{-1}\mu_{2}\right)\). The prediction part relies on the dot product \(wx>C\) where \(w=\varSigma^{-1}\left(\mu_{2}-\mu_{1}\right)\).



\section{Results}

\begin{figure*}[!t]
	\center
	\includegraphics[scale=0.37]{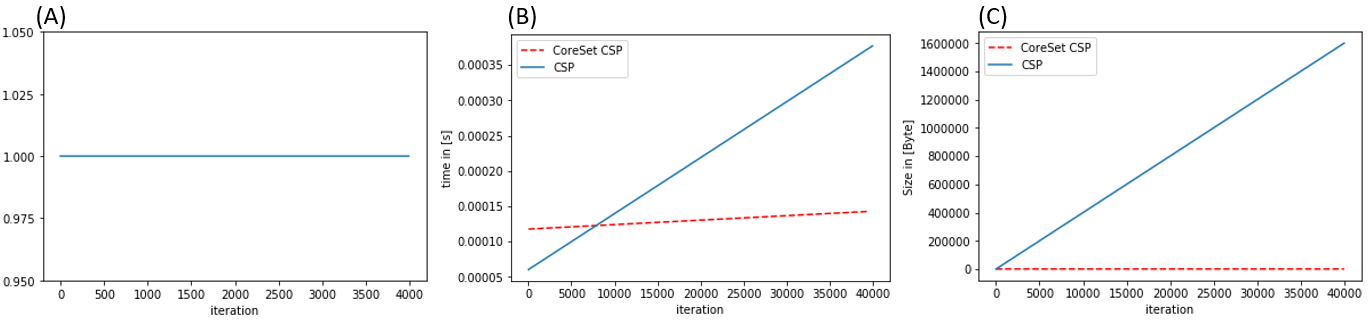}
	\caption{Green line represents traditional CSP algorithm and blue line represents the coreset-based CSP. a) The error between the traditional and coreset-based CSP algorithms, b) The time required for computation as function of number of samples, and c) Memory requirements for both algorithms.} 
	\label{fig:bound}
\end{figure*}

\noindent\textbf{Hardware:} A four core i7 laptop with 8Gb RAM.\\
\textbf{Input Data:} 
For our system evaluation, we use the Motor Imaginary right/left hands task from the EEG Motor Movement/Imaginary dataset that was created and contributed to the Physionet \cite{Goldbergere215} by the developers of BCI2000 instrumentation system \cite{schalk2004bci2000}. The dataset was recorded using 64 electrodes in 10-10 system arrangement, with sampling frequency of 160Hz. The dataset include 109 participants, with about 44 trials (depends on the artifacts rejection process applied) for each MI task.\\
\textbf{Pre-Processing:}
During this step, several sub-routines are applied, as depicted in \ref{fig:scheme} (see second stage). First, an artifacts rejection process is applied in order to (i) remove eye blinks, (ii) detect loosed/noisy electrodes. Second, the signal is band-passed to 0.5-8Hz in order to focus on delta and theta frequencies, which are known to be related with sensi-motor acivitiy \cite{cruikshank2012theta}.
\textbf{Evaluation:} The goal of the experiment was to compare time, memory consumption and accuracy of traditional CSP versus coreset-based CSP, computed at each time sample.
To evaluate our method, we compare the results with traditional CSP-based MI-BCI using the following criterions:
\begin{enumerate}
	\item First we show that the \(w^{T}\) component in both methods reaches the same solution.
	\item Later, a visualization of the 4 best CSP components is compared.
	\item Then we compare the classification accuracy.
	\item Last, we present the time and memory allocation used by both methods.
\end{enumerate}
\hspace*{0.5cm} In order to show that the \(w^{T}\) component reaches to the same solution, we computed the following ratio \(\frac{\left\Vert wX_{1}\right\Vert ^{2}}{\left\Vert wX_{2}\right\Vert ^{2}}/\frac{\left\Vert \tilde{w}X_{1}\right\Vert ^{2}}{\left\Vert \tilde{w}X_{2}\right\Vert ^{2}}\) when \(\tilde{w}\) is calculated using "coreset-based CSP" and \(w\) using the traditional (batch) CSP algorithm. \\
\textbf{Measurements:} The ratio was computed per sample (i.e. to simulate real-time data acquisition conditions), as can be seen in Fig.~\ref{fig:bound}(a), it is noticeable that ratio is stable around the value of 1.

\begin{figure*}[t]
	\center
	\includegraphics[height=2in, width=4in]{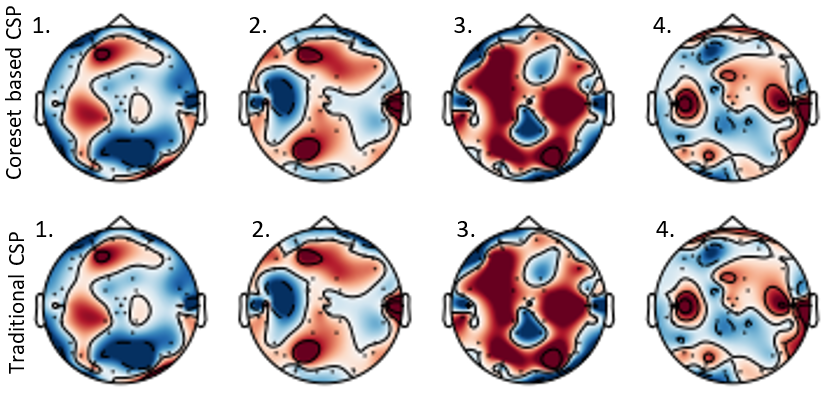}
	\caption{Four best CSP components for discrimination between left and right imaginary hand movement. \newline
		\textbf{Top:} Coreset CSP : \textbf{Bottom:} Traditional CSP}
	\label{fig:topmap}
\end{figure*}

In Fig.~\ref{fig:topmap} we visualize the four largest components computed by each method. The top row shows the coreset-based CSP and at the bottom is the traditional CSP, where highest component is located at the left. It can be seen that the selected CSP weights are the same.

Table~\ref{tbl:res_table} shows the classification accuracy based on a single trial across all 109 participants. The first row presents the average classification accuracy based on CSP computed from EEG data approximated using coreset, and second row show the traditional CSP-based algorithm result. It can be seen that the classification result is not damaged by the coreset approximation of the signal. We cross-validated the data using leave-one-out.
\begin{center} 
	\begin{table} \caption{Classification results - averaged across all participants}
		\label{tbl:res_table}
		\center
		\begin{tabular}{||c c c||}
			\hline
			& mean & std \\ [0.5ex]
			\hline\hline
			Coreset CSP + LDA & 74.9\% & 14.1\% \\ \hline
			Traditional CSP + LDA  & 72.9\% & 13.2\% \\\hline
			
		\end{tabular}
		
	\end{table}
\end{center}
\begin{center} 
	\begin{table} \caption{Classification results - averaged across all participants}
		\label{tbl:c}
		\center
		\begin{tabular}{|c || c | c||}
			\hline
			True Label & Left & Right \\ [0.5ex]\hline
			Left & 0.751 & 0.249 \\ \hline
			Right  & 0.294 & 0.706 \\\hline
			\hline
			Predicted Label & Left & Right\\ \hline
			
		\end{tabular}
		
	\end{table}
	
\end{center}
Table ~\ref{tbl:c} presents the classification results in more details using the confusion matrix showing the type I and type II errors.



Figure ~\ref{fig:bound}(B) presents the computation time  and memory (C) consumption of both methods. It can be seen that the coreset-based algorithm is superior in terms of memory allocation straight from the beginning, and maintains the same level with the time (opposed to the traditional CSP based BCI algorithm). In addition, it can be seen that the coreset-based algorithm is computationally efficient and maintains the same computation time without regard to the number of samples from the signal used.

\section{Conclusions}

We showed that coresets can indeed be used to learn EEG signals in real-time and dynamic data by applying existing algorithms on these coresets. Our theoretical and experimental results demonstrate that this can be done via 64 channels with hundreds of time samples per second, without decreasing the accuracy of the system. Additionally we showed that coreset-based compact representation allows parallel computation and removal (at real time) of bad trials or outliers from the system.

A real time EEG system is valuable for immediate interactive systems such as in neurofeedback settings. Immediate response in a real time EEG system can let the subject learn the system behavior and "teach" himself how to control his thoughts to improve the system's output. Such abilities can shorten the training period of EEG tasks and result with better more adaptive or personalized systems.

Additionally, we show that by using coreset-approximation of the EEG signal, a cheaper (with less memory and computational power) or low powered hardware can be used for training and running the BCI systems.

Full open source is provided~\cite{code} in the hope that it will be used and extended to more coresets and BCI applications in the future.

\section*{References}

\bibliography{sample-bibliography}

\end{document}